\documentclass[runningheads]{svmult}

\usepackage{makeidx}   
\usepackage{graphicx}  
\usepackage{multicol}  
\usepackage{physprbb}  

\def\sch{Schwarzschild}
\begin{document}
\title*{Tomography of the atmosphere of long-period variable stars}
\titlerunning{Journ\'ees Miras 2002, ed. M.O. Mennessier et al., Univ. Montpellier (France)}
%
\author{Alain Jorissen\inst{1}
\and Maya Dedecker\inst{1}
\and Bertrand Plez\inst{2}
\and Denis Gillet\inst{3}
\and Andr\'e Fokin\inst{4}}
\authorrunning{Alain Jorissen et al.}
%
%
\institute{Institut d'Astronomie et d'Astrophysique, 
           Universit\'e Libre de Bruxelles, 
           C.P.\,226, Boulevard du Triomphe,
           B-1050 Bruxelles, Belgium
\and       GRAAL,
           Universit\'e Montpellier II,
           cc072,
           34095 Montpellier cedex 05, France
\and       Observatoire de Haute-Provence,
           04870 Saint-Michel l'Observatoire, France
\and       Institute for Astronomy of the Russia Academy of Sciences,
           48 Pjatnitskaja,
           109017 Moscow, Russia}

\maketitle              

\begin{abstract}
This paper reviews the results obtained so far with the newly developed 
tomographic technique, which probes 
the velocity field across the atmosphere 
of long-period variable (LPV) stars. 
The method 
cross-correlates the optical spectrum with numerical masks constructed from 
synthetic spectra and probing layers of increasing depths. This
technique reveals that the line doubling often observed in LPV stars
around maximum light is the signature of the shock wave
propagating in the atmosphere of these pulsating stars.  
\end{abstract}

\section{Introduction}

It is known since long that the brightness variations of long-period
variable stars (LPVs) go along with  spectral changes such as the doubling
of the absorption lines around maximum  light
\cite{adams41,alvarez00,maehara68}. This line-doubling phenomenon appears 
to be shared by all LPVs \cite{alvarez01b},  and  occurs when 
the shock wave associated with the envelope pulsation propagates through the
line-forming region \cite{alvarez00,schwarzschild52}.
Since it is almost impossible to study individual line profiles in the 
very crowded optical spectra of LPVs, it is necessary to resort to a
technique that correlates the observed spectrum with
a spectral template \cite{alvarez01a,baranne79,queloz95}. The shape of the resulting
cross-correlation function (CCF) thus represents the average 
shape of those lines in the observed spectrum that match the template.
By carefully designing spectral templates collecting all spectral
lines forming at given depths in the atmosphere, it is even possible to 
probe the velocity field in pulsating variables.
This is the general principle guiding
the tomographic\footnote{The word {\it tomography} is used here in its  etymological
sense ({\it `display cuts'}),  which differs somewhat from the broader
sense in use  within the astronomical community (reconstruction of a 
structure using projections taken under different angles).}
method described in this review.
When applied to a temporal sequence of spectra, the method is even able to reveal
in a spectacular way the {\bf outward propagation of the shock wave}, as shown in
Sect.~\ref{Sect:RTCyg} and Figs.~\ref{Fig:RTCyg}--\ref{Fig:RYCep}!

\section{Tomography of the atmosphere}
\label{Sect:tomo}

The study of the velocity field in the atmosphere of LPVs poses a
special challenge, as the spectrum of these stars is extremely 
crowded, particularly in the optical domain. 
The cross-correlation technique provides a powerful tool to overcome this
difficulty. The information relating to the line doubling 
(velocity shift and line shape) is in fact distributed among a large number 
of spectral lines, and can be summed up into an average profile, or more 
precisely into a cross-correlation function (CCF). 
If the correlation of the stellar spectrum with a mask 
involves many lines, it is possible to extract the relevant information from 
very crowded and/or low signal-to-noise spectra. The CCF writes 
\begin{equation}
{\rm CCF}(\Delta\lambda) = \int_{\lambda_1}^{\lambda_2} s(\lambda - \Delta\lambda)\;
m(\lambda)\; {\rm d}\lambda,
\end{equation}
where $s(\lambda)$ is the observed spectrum, $m(\lambda)$ is the template (a binary
template has been adopted, being  0 around spectral lines of interest, and 1
elsewhere),
and $\lambda_1$ and $\lambda_2$ are the boundaries of the spectral range covered by the
observed spectrum. The radial velocity  $V_r$ is then obtained from the wavelength
shift $\Delta\lambda$(min CCF) where the CCF is minimum:
\begin{equation}
V_r = \frac{\Delta\lambda({\rm min \; CCF}) }  {0.5\; (\lambda_1 + \lambda_2)} \;c,
\end{equation}
where $c$ is the speed of light in vacuum.
We refer the reader to reference
\cite{baranne79} for a detailed  description of the CCF mathematical properties.

The tomographic method rests on our ability to construct reliable
synthetic spectra of 
late-type giant stars \cite{plez92a,plez92b,plez99}, and 
from those, 
to identify the depth of formation of any given spectral line. 
Rigorously, the contribution function (CF) to the
flux depression \cite{albrow96} should be used to evaluate the
geometrical depth at which a line forms.  However, it
would be a formidable task to compute the CF for each line appearing
in the optical spectrum of LPV stars. For the sake of simplicity,
the `depth function' $x = x(\lambda)$ is used instead, which  
provides the geometrical depth corresponding to monochromatic 
optical depth $\tau_\lambda = 2/3$\ at the  considered wavelength $\lambda$. 
This function expresses the
depth from which the emergent flux arises, in the Eddington-Barbier
approximation, and should not 
differ much from the average depth of formation for sufficiently strong lines \cite{magain86}.

\begin{figure}
\includegraphics[angle=90,width=1.0\textwidth]{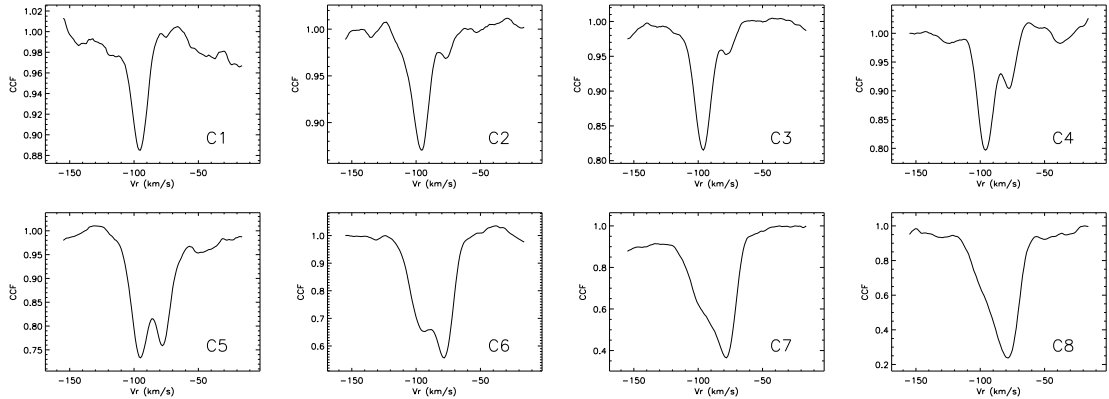}
\vspace{-6cm}
\caption[]{
The CCFs of the Mira Z Oph at phase
0.08 obtained with the tomographic masks.
 Note how the shape of the CCFs evolve from the innermost layer
(involving ascending matter only, hence C1 exhibits a single blue peak) 
to the outermost layer (involving mostly matter falling in, hence C8
exibits predominantly a red peak). This spatial sequence of line
doubling  reflects the presence of a shock wave in the
line-forming region, with
the shock front being centered on the layer probed by the mask C5}
\label{Fig:ZOph2}
\end{figure}

\begin{figure}
\includegraphics[angle=90,width=1.0\textwidth]{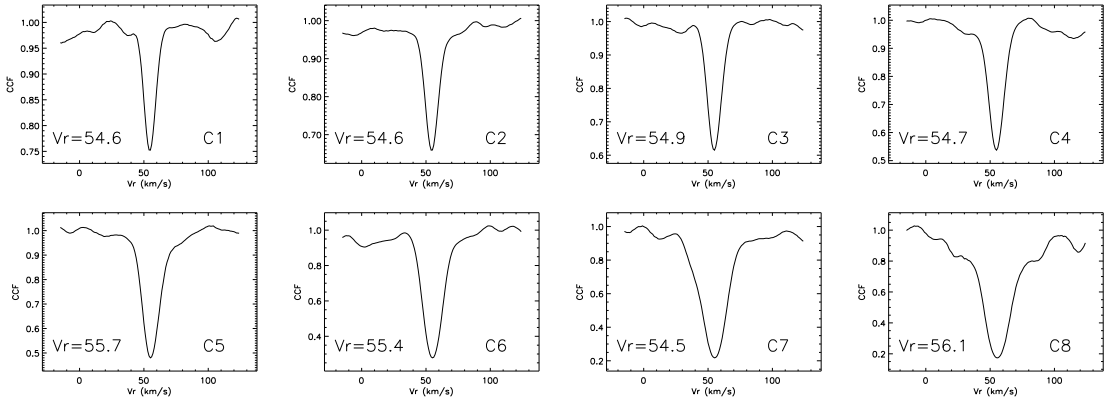}
\vspace{-6cm}
\caption[]{Same as Fig.~\protect\ref{Fig:ZOph2} 
for the non-Mira star $\mu$ Gem.
Note how the CCFs broaden and become more contrasted in the upper atmospheric layers
(C8 is the outermost mask), reflecting the fact that the strongest spectral lines
form in the outermost layers. The radial velocity (in km/s) derived from each
CCF is given in each panel.  
}
\label{Fig:constant}
\end{figure}

Different masks $C_i$ are then constructed 
from the collection of $N_i$ lines $\lambda_{i,j} (1 \le j \le N_i)$ such that 
$x_i \le x(\lambda_{i,j}) < x_{i+1} = x_i + \Delta x$, where $\Delta x$ is 
some constant optimized to keep enough lines in any given mask without 
losing too much resolution in terms of geometrical depth. A natural
limit to the resolution that can be achieved on the geometrical depth
is provided by the width of the CF \cite{alvarez01a}, and according to 
the Nyquist-Shannon theorem of elementary signal theory, there is no
advantage in taking $\Delta x$ smaller than half the CF width of typical
lines probed by the mask.
Each mask $C_i$ 
should then probe lines forming at (geometrical) depths in the range 
$x_i, x_i + \Delta x$ in the atmosphere. 
These masks are used as templates to correlate with the observed 
spectra of the LPV stars. This procedure should provide the velocity field 
as a function of depth in the atmosphere (Fig.~\ref{Fig:ZOph2}), 
whereas for static atmospheres, all CCFs
should yield the same radial velocity (Fig.~\ref{Fig:constant}).

A more detailed
description of the method can be found in \cite{alvarez01a}, while specific
tomographic masks are available at
{\it http://www-astro.ulb.ac.be/Html/home.html\#tomography}.
The set of tomographic masks used in this paper was constructed from a synthetic
spectrum at $T_{\rm eff}$ = 3500~K and $\log g = 0.9$ (see reference
\cite{alvarez01a} for details). The properties of the masks (in terms
of numbers of lines and depths probed) are listed 
in Table~\ref{Tab:masks}. The depth $x$ is in fact expressed in terms of 
a reference optical depth $x \equiv \log \tau_0$, where $\tau_0$ is the
optical depth at the reference wavelength of 1.2 $\mu$m. There is a
one-to-one correspondence between $\log \tau_0$ and the geometrical depth.

\begin{table}
\caption[]{The synthetic templates constructed from the model at
  3500~K and $\log g = 0.9$.}
\begin{center}
\tabcolsep 10pt
\begin{tabular}{lccc}
\hline\noalign{\smallskip}
Mask \# & depth probed & \multicolumn{1}{c}{\# of lines per mask} \\
        & $x \equiv \log \tau_{1.2 \mu m}$\\
$C_i$     & $x_i, x_i + \Delta x$ & $N_i$ \\ 
\hline\noalign{\smallskip}

$C_1$ (innermost) & $-2.00, -2.75$ & 777  \\ 
$C_2$             & $-2.75, -3.50$ & 610  \\
$C_3$             & $-3.50, -4.25$ & 433  \\
$C_4$             & $-4.25, -5.00$ & 321  \\
$C_5$             & $-5.00, -5.75$ & 168  \\
$C_6$             & $-5.75, -6.50$ & 167  \\
$C_7$             & $-6.50, -7.25$ &  94  \\
$C_8$ (outermost) & $-7.25, -8.00$ &  46  \\
\noalign{\smallskip}
\hline
\end{tabular}
\end{center}
\label{Tab:masks}
\end{table}

\begin{figure}
\begin{center}
\vspace{-5cm}
\includegraphics[angle=0,width=1.0\textwidth]{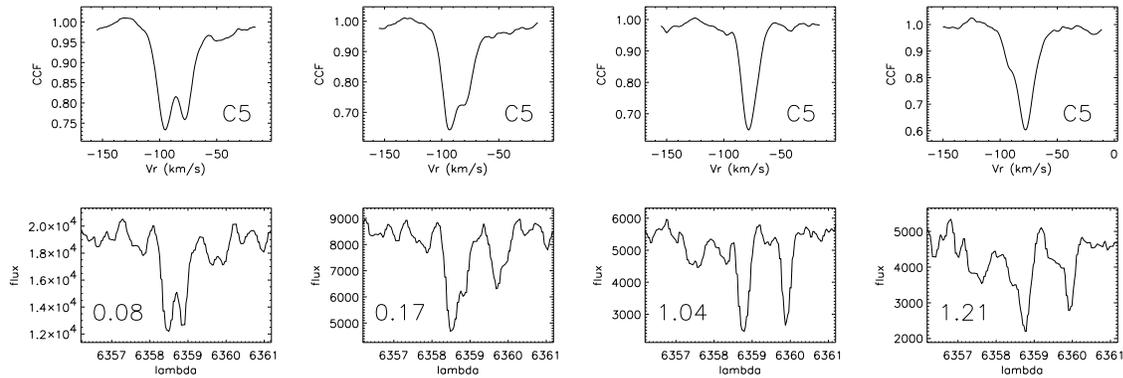}
\end{center}
\caption[]{Comparison of the $\lambda$ 6358.69 FeI line as seen in the spectrum of Z
Oph at phases 0.08, 0.17, 1.04 and 1.21 (bottom row, from left to right), with the
CCF obtained with the tomographic mask C5 (upper row), 
to which belongs the $\lambda$ 6358.69 FeI
line. Z Oph is warm enough at maximum light (K3) for its spectrum
to be not too crowded so that clean, almost unblended,  lines as the
one displayed here may be isolated}
\label{Fig:ZOph}
\end{figure}

\newpage

\begin{figure}
\begin{center}
\includegraphics[height=6cm]{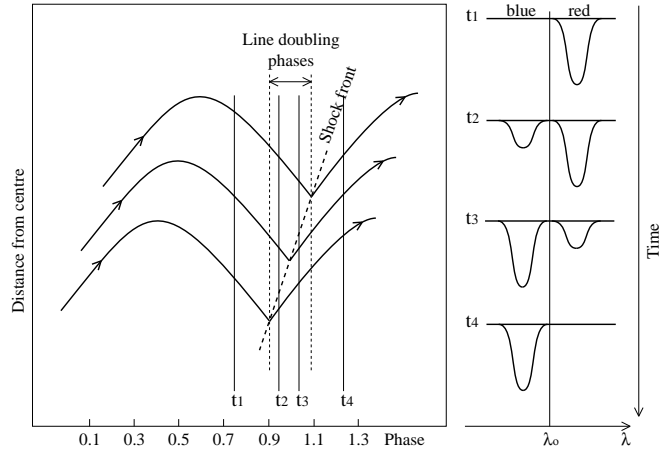}
\end{center}
\caption[]{The Schwarzschild scenario: temporal sequence followed by the 
   intensity of the red and blue components of absorption lines close to 
   maximum light, when the shock wave propagates through the photosphere}
   \label{Fig:Schwarzschild}
\label{Fig:schwarz}
\end{figure}

An {\it a posteriori} validation of the
method is provided by the fact that (i) the CCFs obtained with the tomographic
masks for the non-pulsating M giant $\mu$~Gem (Lb variable with a variability
range of only 0.11~mag in the Hipparcos catalogue) are all single-lined and
yield almost identical radial velocities, as expected (Fig.~\ref{Fig:constant}), and
(ii) the asymmetric CCFs obtained for the warm Mira variable Z Oph reflect the shape
of the individual lines probed by any given mask (Fig.~\ref{Fig:ZOph}). The
asymmetric, and sometime even double-peaked, CCFs  observed for Mira variables  thus
correctly represent the underlying spectral lines, and are not artifacts of the
correlation process. Moreover, around maximum light, the shape of CCFs 
in the successive
tomographic masks follows a definite sequence: the CCF in the
deepest layer exhibits a single blue component whereas a single red
component
is observed in the outermost CCF, and double-peaked CCFs
are observed in between (Fig.~\ref{Fig:ZOph2}). This sequence, which is observed in
almost all Mira around maximum light, is consistent with the \sch\ scenario
describing the effects of the passage of a shock wave through the atmosphere of a
pulsating star (Sect.~\ref{Sect:schw}).

\section{The \sch\ scenario}
\label{Sect:schw}

The passage of a shock wave through a stellar atmosphere is expected to give rise to
a very specific temporal evolution of the spectral line shape
(Fig.~\ref{Fig:schwarz}), as first noted by Schwarzschild \cite{schwarzschild52} in 
relation with W Vir Cepheids. When the shock wave penetrates the line-forming
region, the velocity discontinuity associated with the shock gives rise to the
doubling of the spectral line: the red component of the line is formed in the 
matter falling in above the shock, whereas the blue component arises from the
ascending matter lying behind the shock.
The intensity of the blue and red components of 
a double line observed around maximum light in a LPV star should follow 
the temporal sequence  illustrated in Fig.~\ref{Fig:Schwarzschild}. 
This is indeed the case, as revealed by the evolution of the CCFs
of the Mira variables RT~Cyg and RY~Cep around maximum light
(Sect.~\ref{Sect:RTCyg}). But the tomographic method
described in Sect.~\ref{Sect:tomo} allows us to go one step further, 
by showing that the Schwarzschild scenario also reveals itself in
terms of a definite 
{\it spatial} sequence of profiles (Fig.~\ref{Fig:ZOph2}). By
combining those spatial sequences obtained at successive phases in the
light cycle, the tomographic method reveals that   
{\it line-doubling appears later  in upper layers of the Mira atmosphere, thus
translating the upward motion of the shock wave} (Sect.~\ref{Sect:RTCyg}).

\section{Application to the Mira variables RT Cyg and RY Cep}
\label{Sect:RTCyg}

A long-term monitoring of the  
Mira stars RT\,Cyg ($P = 190$~d; $6.0 \le V
\le 13.1$) and RY~Cep ($P = 145$~d; $8.6 \le V
\le 13.6$) has been performed  with 
the fibre-fed echelle spectrograph ELODIE \cite{baranne96}. 
The spectrograph  ELODIE is
mounted on  the 1.93-m telescope of the Observatoire de Haute Provence (France), and 
covers the full range from 3906~\AA\ to 6811~\AA\ in one exposure at a 
resolving power of 42\,000.

\begin{figure}
\hspace*{2cm}C3\hspace{1.5cm} C4 \hspace{1.5cm} C5 \hspace{1.5cm} C7
\hspace{1.2cm} phase
\begin{center}
\vspace{-5mm}
\includegraphics[angle=90,height=0.7\textheight]{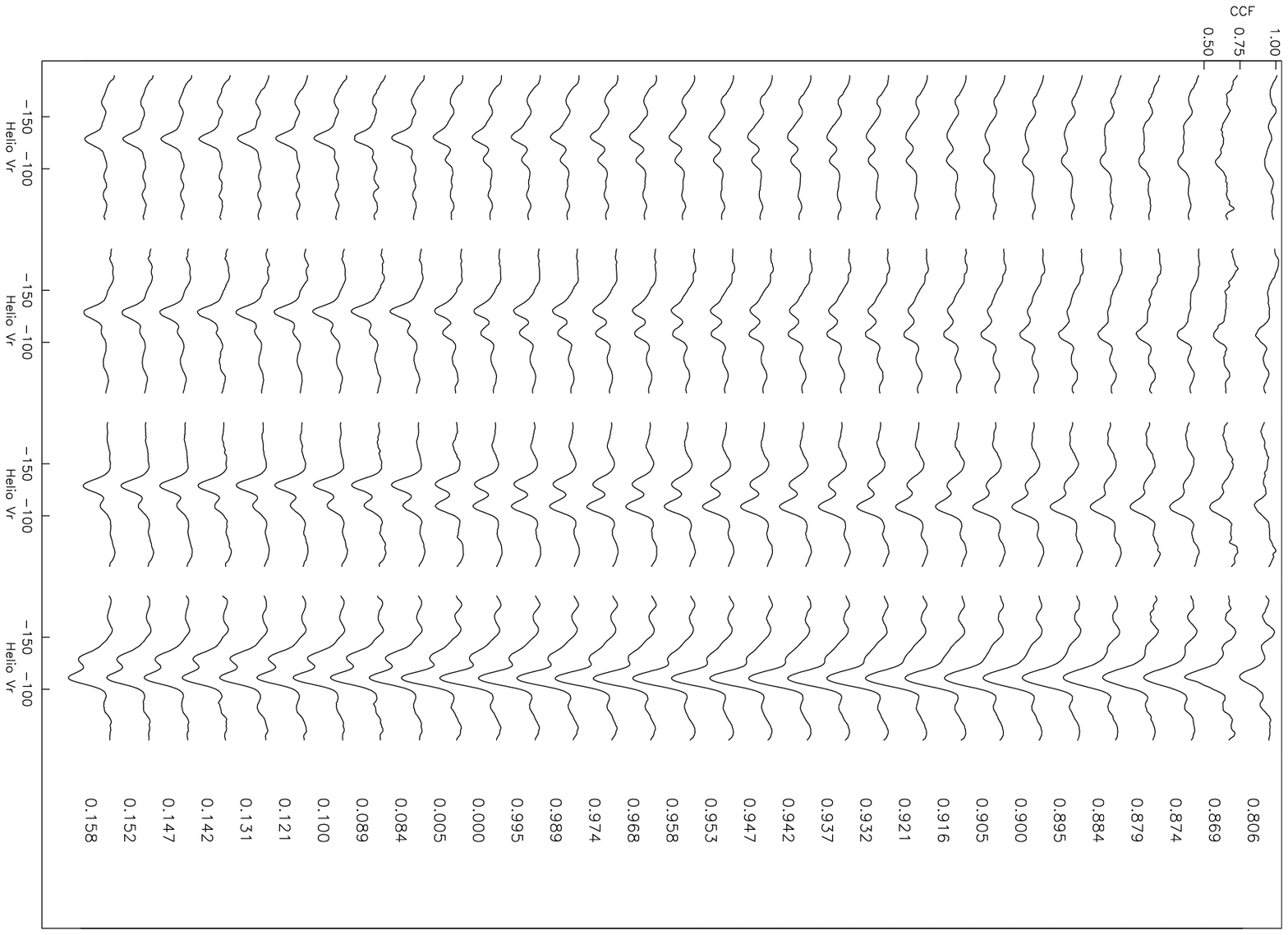}
\mbox{}
\hspace{-2.5cm}
\includegraphics[height=0.2\textheight,width=0.8\textwidth]{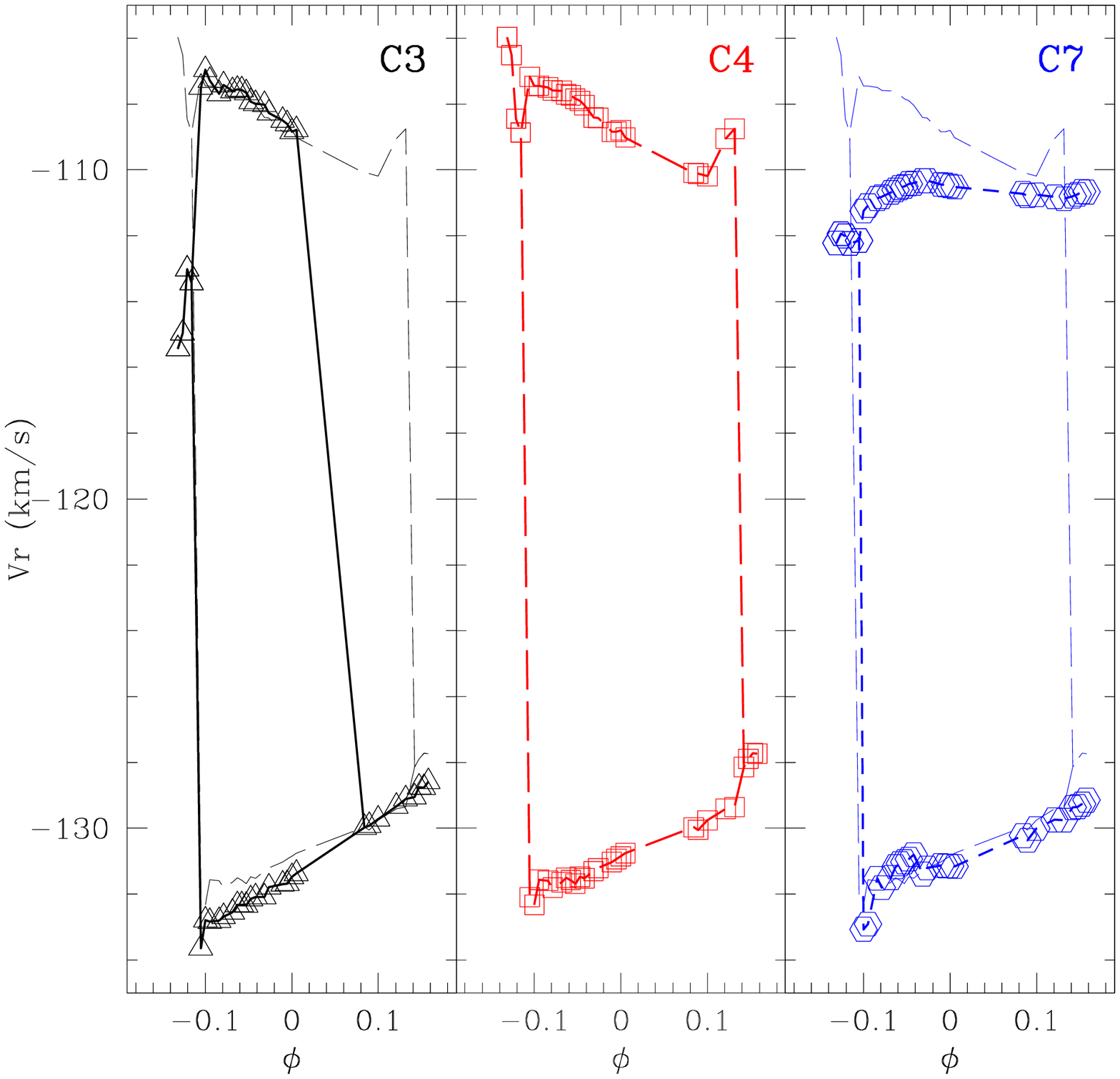}
\end{center}
\caption[]{Top panel: Sequence of cross-correlation profiles of RT\,Cyg 
 obtained with the tomographic masks C3, C4, C5 and C7 (from left to right) in
August-September 1999, around maximum light. 
The labels beside each CCF refer to the
 corresponding visual phase based on the AAVSO ephemeris \cite{mattei99}\\
Bottom panels: Radial velocity curves obtained with the
  tomographic masks C3 (leftmost panel - solid line - triangles), C4
  (middle panel - long-dashed line 
  - squares), and C7 (rightmost panel - short-dashed line -
  circles). The C4 radial-velocity curve (long-dashed line) 
has been duplicated in the
  left- and rightmost panels to allow an easy intercomparison 
}
\label{Fig:RTCyg}
\end{figure}

\begin{figure}
\hspace*{2cm}C3\hspace{1.5cm} C4 \hspace{1.5cm} C5 \hspace{1.5cm} C6
\hspace{1.2cm} phase
\begin{center}
\vspace{-1cm}
\includegraphics[angle=90,height=0.7\textheight]{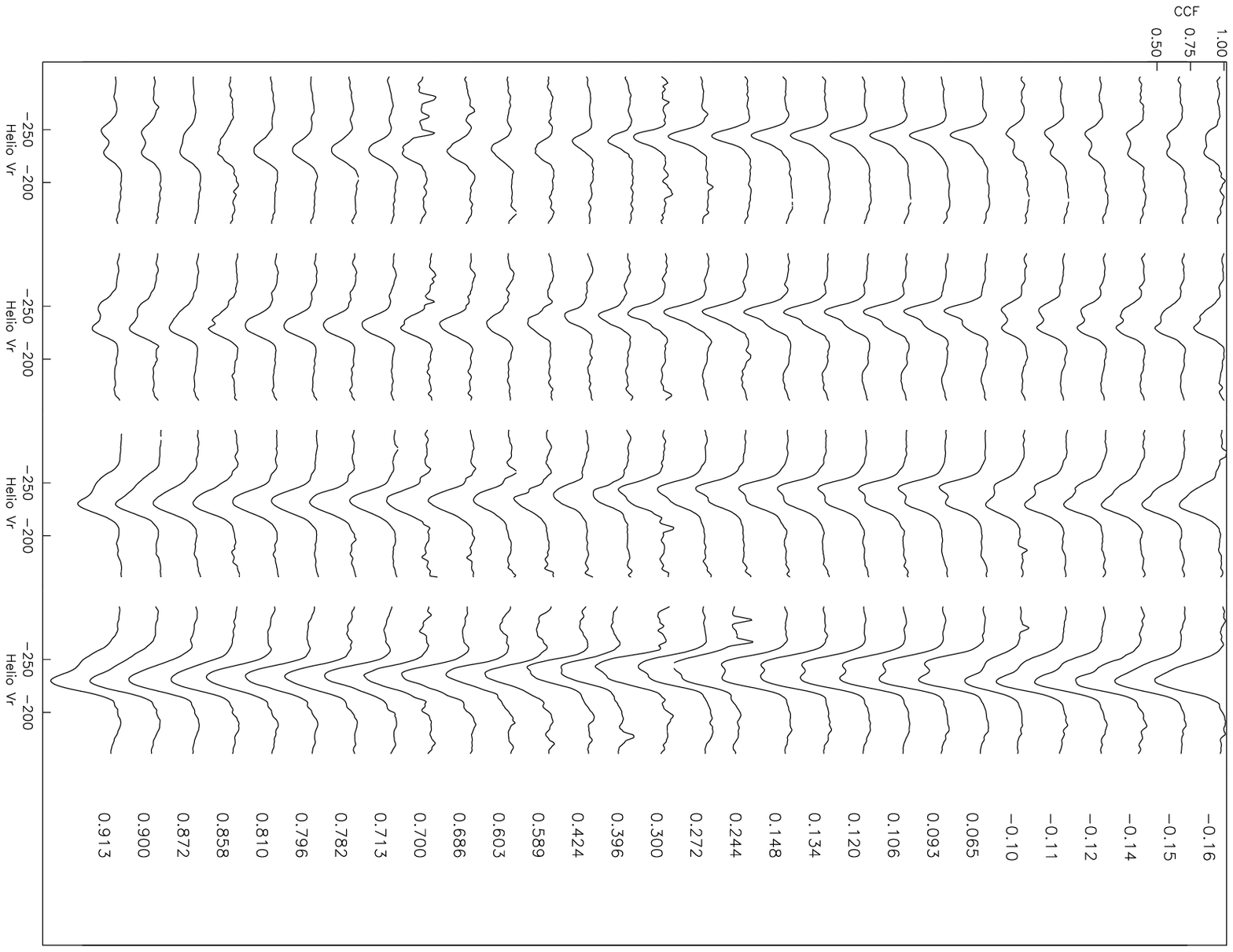}
\mbox{}
\hspace{-2.5cm}
\includegraphics[height=0.2\textheight,width=0.8\textwidth]{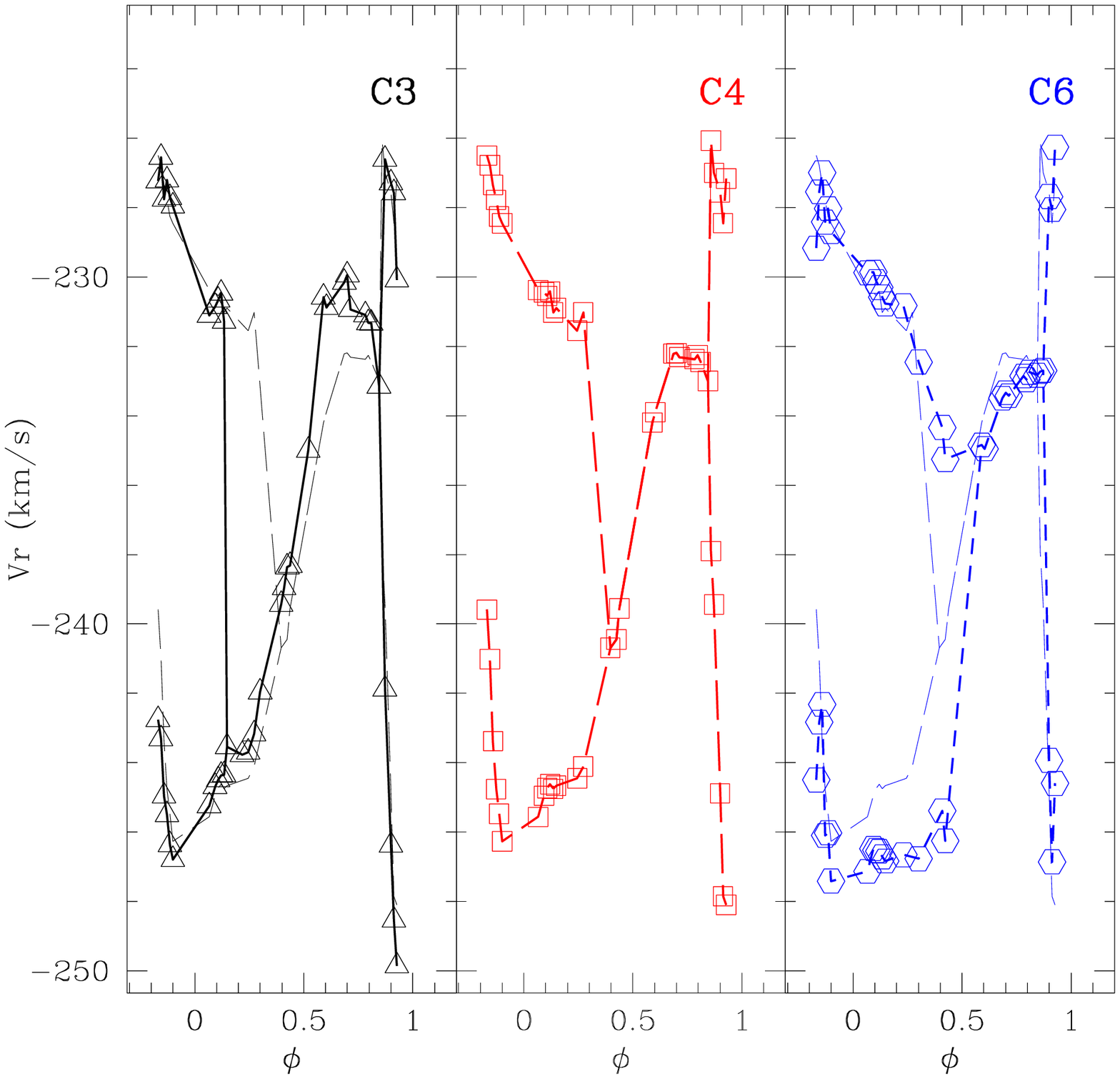}
\end{center}
\caption[]{Top panel: Same as Fig.~\protect\ref{Fig:RTCyg} for RY Cep with the
  tomographic masks C3 (inner), C4, C5 and C6 (outer), from left to right.
The labels beside each CCF refer to the
 corresponding visual phase based on the AFOEV monitoring.\\
Bottom panels: Radial velocity curves obtained with the
  tomographic masks C3 (leftmost panel - solid line - triangles), C4
  (middle panel - long-dashed line 
  - squares), and C6 (rightmost panel - short-dashed line -
  circles). The C4 radial-velocity curve (long-dashed line) 
has been duplicated in the
  left- and rightmost panels to allow an easy intercomparison 
}
\label{Fig:RYCep}
\end{figure}

For RT~Cyg, a monitoring in August-September 1999 covered phases
$-0.20$ to 0.16 around maximum light, with 32 spectra obtained during this phase
range (corresponding to an average resolution of $\Delta\phi = 0.01$;
Fig.~\ref{Fig:RTCyg}).   For RY~Cep, the monitoring covered a {\it full} light cycle
extending from August 2001 (phase -0.17) to February 2002 (phase 0.92), and 40
spectra were obtained (Fig.~\ref{Fig:RYCep}).

The sequences of CCFs presented in Figs.~\ref{Fig:RTCyg} and \ref{Fig:RYCep} clearly
obey the \sch\ temporal evolution, with a single red component transforming
progressively into a single blue component around maximum light. 
There are moreover clear phase lags
between the different layers,  this
transformation occurring at later phases in outer layers. 
Thus, the \sch\ scenario holds 
for Mira variables, and this conclusion  definitely points towards the velocity
stratification associated with the shock wave as the cause of 
the double absorption lines observed in Mira variables, as
opposed to complex radiative processes (e.g., radiative 
release of thermal energy into the post-shock layer or temperature inversion) 
\cite{gillet85,karp75}. 

Radial velocities have been extracted from the CCFs of 
Figs.~\ref{Fig:RTCyg} and \ref{Fig:RYCep} (top panels) to yield the
curves displayed in the bottom panels of the same figures.
We defer the detailed comparison between the radial-velocity curves of
RY~Cep and RT~Cyg, and their interpretation, to a forthcoming paper.  
We just stress here noteworthy features on
the radial-velocity curve of RY Cep: 
(i) the red peaks (corresponding to matter falling in) 
are observed at the same velocity in
all three masks; they disappear at increasingly later phases as one
considers layers closer and closer to the surface, since the outermost 
layers are the last ones to be penetrated by the shock wave
which suppresses the infalling 
component (i.e., red peak);   
(ii) the maximum outward velocity
is the same for the three masks ($\sim -247$ km/s), 
and is reached around phase $-0.1$. After that, matter will decelerate 
in the innermost layers first, as they lose the impetus provided by
the shock wave which is moving away.

\section{Conclusion and perspectives}

The tomographic technique presented in this paper 
opens new perspectives for the study of the 
dynamics of LPV stars, as it allows 
to visualize the outward motion of the shock 
wave in the atmosphere. At this stage, however, the results remain
qualitative as it is difficult to quantify the geometrical depth of
the layers probed by the different masks. It is hoped that, in the future, 
this method may be combined with interferometric imaging in the
different masks to derive the corresponding radii. This would 
give direct access to the velocity of the shock, for instance.\\

\noindent
{\small
{\it Acknowledgements}. We thank the Observing Program Committee of the
Observatoire de Haute-Provence for a generous allocation of telescope time.
The AFOEV is warmly thanked for positively answering  our request to monitor 
RY Cep during the time of the spectroscopic monitoring.
 A.J. is
Research Associate from the {\it Fonds National de la Recherche Scientifique}
(Belgium). Financial support has been received from the {\it Communaut\'e Fran\c caise
  de Belgique} in the framework of a TOURNESOL programme.}

\end{document}